\documentclass[12pt]{article}
\usepackage{amsfonts,graphicx,amsmath,amsthm,amssymb,upgreek,epsfig}
\usepackage{float}
\allowdisplaybreaks

\begin{document}

\title{\bf Study of Anisotropic Compact Stars in $f(\mathcal{R},\mathcal{T},\mathcal{R}_{\chi\xi}\mathcal{T}^{\chi\xi})$ Gravity}
\author{M. Sharif \thanks{msharif.math@pu.edu.pk}~ and T. Naseer \thanks{tayyabnaseer48@yahoo.com;
tayyab.naseer@math.uol.edu.pk}\\
Department of Mathematics, University of the Punjab,\\
Quaid-i-Azam Campus, Lahore-54590, Pakistan.}

\date{}
\maketitle

\begin{abstract}
This paper aims to examine the composition of various spherically
symmetric star models which are coupled with anisotropic
configuration in $f(\mathcal{R},\mathcal{T},\mathcal{Q})$ gravity,
where $\mathcal{Q}=\mathcal{R}_{\chi\xi}\mathcal{T}^{\chi\xi}$. We
discuss the physical features of compact objects by employing bag
model equation of state and construct the modified field equations
in terms of Krori-Barua ansatz involving unknowns ($A,B,C$). The
observational data of 4U 1820-30,~Vela X-I,~SAX J 1808.4-3658,~RXJ
1856-37 and Her X-I is used to calculate these unknowns and bag
constant $\mathfrak{B_c}$. Further, we observe the behavior of
energy density, radial and tangential pressure as well as anisotropy
through graphical interpretation for a viable model
$\mathcal{R}+\varrho\mathcal{Q}$ of this gravity. For a particular
value of the coupling constant $\varrho$, we study the behavior of
mass, compactness, redshift and the energy bounds. The stability of
the considered stars is also checked by using two criteria. We
conclude that our developed structure in this gravity is in
well-agreement with all the physical requirements.
\end{abstract}
{\bf Keywords:}
$f(\mathcal{R},\mathcal{T},\mathcal{R}_{\chi\xi}\mathcal{T}^{\chi\xi})$
gravity; Anisotropy; Compact stars. \\
{\bf PACS:} 04.20.Jb; 98.80.Jk; 03.50.De.

\section{Introduction}

General Relativity (GR) has accomplished remarkable results in
resolving numerous hidden ingredients of the universe, however it is
not satisfactory enough to scrutinize the cosmos at large scale.
Many other theories alternative to GR are therefore established as
the efficient approaches to tackle the challenging mysteries such as
dark matter and cosmic accelerated expansion. Such expansion
guarantees the existence of an obscure form of force with large
negative pressure, known as dark energy. Thus the modified
gravitational theories have been labeled extremely significant to
unveil the enigmatic features of our universe. The first ever
modification to GR is $f(\mathcal{R})$ theory which involves higher
order curvature terms due to the insertion of generic function of
the Ricci scalar in place of $\mathcal{R}$ in an Einstein-Hilbert
action. The physical feasibility of different stellar structures has
been discussed by utilizing multiple techniques in this theory
\cite{2}-\cite{2b}. Capozziello \emph{et al.} \cite{8} studied
various mathematical models and analyzed their stability through the
Lan\'{e}-Emden equation in $f(\mathcal{R})$ theory. Numerous
research \cite{9}-\cite{9d} has been done in this context to
investigate the composition and evolution of astrophysical bodies.

The notion of matter-geometry coupling was initially presented by
Bertolami \emph{et al.} \cite{10} to explore more interesting
features of our universe. They considered the matter Lagrangian as a
function of $\mathcal{R}$ and $\mathcal{L}_{m}$ to study the
influence of coupling on stellar objects in $f(\mathcal{R})$
gravity. Such interaction between geometry and matter distribution
encouraged many researchers to focus on the universal accelerating
expansion. Recently, different modified theories have been proposed
which interlinked the matter and geometry of massive structures at
the action level. Harko \emph{et al.} \cite{20} extended the
$f(\mathcal{R})$ theory to $f(\mathcal{R},\mathcal{T})$ in which
$\mathcal{T}$ indicates trace of the energy-momentum tensor
$(\mathrm{EMT})$. Note that the dependence from $\mathcal{T}$ may be
induced by exotic imperfect fluids or quantum effects. Since in the
present model the covariant divergence of the $\mathrm{EMT}$ is
non-zero, the motion of massive test particles is non-geodesic and
an extra force, orthogonal to the four-velocity is always present
due to the coupling between matter and geometry. This force also
helps to elucidate the galactic rotation curves. The fascinating
results provided by this theory has prompted numerous scientists to
study the astrophysical structures \cite{21}-\cite{21d}. Soon after
this, Haghani \emph{et al.} \cite{22} proposed the extension of
$f(\mathcal{R},\mathcal{T})$ gravity by considering a more
complicated functional of the form
$f(\mathcal{R},\mathcal{T},\mathcal{Q})$ in which the factor
$\mathcal{Q}\equiv \mathcal{R}_{\chi\xi}\mathcal{T}^{\chi\xi}$
guarantees the presence of strong non-minimal coupling even in the
case of traceless $\mathrm{EMT}$. They also analyzed the impact of
term $\mathcal{Q}$ on the feasibility of various models and
concluded that the Lagrange multiplier approach provides conserved
equations of motion in this theory. Sharif and Zubair considered two
particular models in this scenario and calculated their energy
bounds as well as the conditions for Dolgov-Kawasaki instability
\cite{22a}. They also studied the black hole laws of thermodynamics
with different choices of the matter Lagrangian \cite{22b}.

Odintsov and S\'{a}ez-G\'{o}mez \cite{23} calculated the solution of
complex field equations for various models through numerical methods
in $f(\mathcal{R},\mathcal{T},\mathcal{Q})$ gravity and stressed
some serious difficulties associated with the matter instability.
Ayuso \emph{et al.} \cite{24} studied celestial objects and adopted
some scalar and vector fields to obtain the stability conditions for
those structures in this theory. They deduced that the matter
instability must appear in the case of vector field. Baffou \emph{et
al.} \cite{25} analyzed the viability of the solution of modified
equations of motion through the incorporation of perturbation
functions. Sharif and Waseem \cite{25a,25b} have done a
comprehensive analysis on physical features of different neutron
star candidates coupled with isotropic as well as anisotropic
configuration in this gravity. Yousaf \emph{et al.}
\cite{26}-\cite{26e} found the effective structure scalars in
$f(\mathcal{R},\mathcal{T},\mathcal{Q})$ scenario to study the
evolution of spherical and cylindrical static as well as non-static
structures. In this scenario, we examined some physical
characteristics of charged/uncharged compact structures through
gravitational decoupling \cite{26f,26g}.

Stars are identified as astronomical objects which play a
fundamental role in the formation of galaxies in our universe.
Numerous astrophysicists concentrated on the study of their
structure and evolutionary phases. The inward gravitational force
induced due to the mass of a star is counterbalanced by outward
pressure which is generated as a result of nuclear reactions
occurring in the core of stars. When pressure is no longer enough to
resist the attractive force of gravity, there occurs a gravitational
collapse resulting in the death of stellar object due to which
different celestial objects such as white dwarfs, neutron stars and
black holes come into existence. The investigation of such compact
stars led many astronomers to examine their diverse properties.
Among these massive objects, neutron stars have attracted
considerable interest owing to the composition and fascinating
features of their structure. Neutrons produce degeneracy pressure
which counterbalances the gravitational pull and helps to keep them
in hydrostatic equilibrium. In between neutron star and black hole,
there is a quark star which is highly dense structure consisting of
up, down and strange quark matter. Many researchers
\cite{27}-\cite{28c} analyzed the inner formation of these
speculative objects.

The anisotropic configured bodies play a decisive role in the study
of their structural characteristics. The interior of compact stars
should possess anisotropic pressure as they encompass the density
much higher than nuclear density \cite{29}. Herrera and Santos
\cite{30} examined the persuasive causes and impact of anisotropy on
massive structures. Harko and Mak \cite{31} considered a particular
form of anisotropic factor and calculated interior solutions for
static relativistic objects. Hossein \emph{et al.} \cite{32} studied
the effects of cosmological constant $\Lambda$ on massive
anisotropic structures and examined their stability. Kalam \emph{et
al.} \cite{32a} analyzed the validity of energy conditions and
stability of different anisotropic neutron stars. Paul and Deb
\cite{33} investigated the anisotropic configured compact objects
and developed physically feasible solutions.

It is anticipated that the MIT bag model equation of state (EoS)
helps to express the interior configuration of quark bodies
\cite{27}. Of particular interest, the compactness of celestial
structures such as 4U 1820-30, 4U 1728-34, SAX J 1808.4-3658, Her
X-1, RXJ 185635-3754 and PSR 0943+10, etc., cannot be explained by
the neutron star EoS, while MIT bag model (strange quark matter EoS)
\cite{33a} expresses their compactness successfully. The discrepancy
between true and false vacuum can be calculated through the bag
constant $\mathfrak{B_c}$ appearing in the bag model EoS, the
increment of whose value causes the quark pressure to decrease.
Several investigators \cite{33b,34} utilized the MIT bag model EoS
to predict the quarks' inner fluid distribution. Demorest \emph{et
al.} \cite{34a} calculated the mass of a particular quark star (PSR
J1614-2230) and concluded that only the MIT bag model EoS supports
such heavily objects. Rahaman \emph{et al.} \cite{35} examined some
physical characteristics of a strange star having radius $9.9$ km
and calculated the mass of different stars through an interpolating
function. A hybrid star model has been presented by Bhar \cite{36}
through Krori-Barua ansatz and the calculated mass function was
found to be compatible with the observational data. Arba\~{n}il and
Malheiro \cite{37} determined the numerical solution of the
hydrostatic equilibrium condition, radial perturbation as well as
MIT bag model to study the effects of anisotropy on the feasibility
of compact stars. Deb \emph{et al.} \cite{37a,37b} studied
charged/uncharged strange stars, constructed the corresponding
non-singular anisotropic solutions by employing the same EoS and
checked their viability through graphical observation. Sharif and
his collaborators \cite{38}-\cite{38f} determined anisotropic
solutions corresponding to different star candidates with the help
of MIT bag model.

This paper analyzes the influence of anisotropy on different quark
stars in view of the Krori-Barua solution for a particular model in
$f(\mathcal{R},\mathcal{T},\mathcal{Q})$ scenario. The paper is
structured as follows. The formulation of modified field equations
in terms of MIT bag model and Krori-Barua ansatz is presented in the
next section. In section 3, we use the junction conditions at the
boundary to calculate Krori-Barua constants. The graphical behavior
of various physical features of all the considered stars is analyzed
in section 4. Lastly, section 5 provides the concluding remarks.

\section{The $f(\mathcal{R},\mathcal{T},\mathcal{Q})$ Gravity}

The modification of Einstein-Hilbert action (with $\kappa=8\pi$)
involving complex analytical functional $f$ is defined as \cite{23}
\begin{equation}\label{g1}
S_{f(\mathcal{R},\mathcal{T},\mathcal{R}_{\chi\xi}\mathcal{T}^{\chi\xi})}=\int\sqrt{-g}
\left[\frac{f(\mathcal{R},\mathcal{T},\mathcal{R}_{\chi\xi}\mathcal{T}^{\chi\xi})}{16\pi}
+\mathcal{L}_{m}\right]d^{4}x,
\end{equation}
where $\mathcal{L}_{m}$ represents the Lagrangian density of fluid
configuration. Corresponding to the action \eqref{g1}, the field
equations take the form as
\begin{equation}\label{g2}
\mathcal{G}_{\chi\xi}=\mathcal{T}_{\chi\xi}^{(eff)}=-\frac{8\pi}{\mathcal{L}_{m}f_{\mathcal{Q}}
-f_{\mathcal{R}}}\mathcal{T}_{\chi\xi}^{(m)}+\mathcal{T}_{\chi\xi}^{(\mathcal{D})}.
\end{equation}
The term $\mathcal{G}_{\chi\xi}$ expresses the geometric structure
of the celestial bodies whereas $\mathcal{T}_{\chi\xi}^{(eff)}$ is
identified as the $\mathrm{EMT}$ in
$f(\mathcal{R},\mathcal{T},\mathcal{Q})$ gravity which involves
physical variables along with modified corrections. In this
scenario, the sector $\mathcal{T}_{\chi\xi}^{(D)}$ appearing due to
modified gravity has the form
\begin{eqnarray}\nonumber
\mathcal{T}_{\chi\xi}^{(\mathcal{D})}&=&-\frac{1}{\mathcal{L}_{m}f_{\mathcal{Q}}-f_{\mathcal{R}}}
\left[\left(f_{\mathcal{T}}+\frac{1}{2}\mathcal{R}f_{\mathcal{Q}}\right)\mathcal{T}_{\chi\xi}^{(m)}
+\left\{\frac{\mathcal{R}}{2}(\frac{f}{\mathcal{R}}-f_{\mathcal{R}})-\mathcal{L}_{m}f_{\mathcal{T}}\right.\right.\\\nonumber
&-&\left.\frac{1}{2}\nabla_{\lambda}\nabla_{\eta}(f_{\mathcal{Q}}\mathcal{T}^{\lambda\eta})\right\}g_{\chi\xi}
-\frac{1}{2}\Box(f_{\mathcal{Q}}\mathcal{T}_{\chi\xi})-(g_{\chi\xi}\Box-
\nabla_{\chi}\nabla_{\xi})f_{\mathcal{R}}\\\label{g4}
&-&2f_{\mathcal{Q}}\mathcal{R}_{\lambda(\chi}\mathcal{T}_{\xi)}^{\lambda}+\nabla_{\lambda}\nabla_{(\chi}[\mathcal{T}_{\xi)}^{\lambda}f_{\mathcal{Q}}]
+2(f_{\mathcal{Q}}\mathcal{R}^{\lambda\eta}+\left.f_{\mathcal{T}}g^{\lambda\eta})\frac{\partial^2\mathcal{L}_{m}}
{\partial g^{\chi\xi}\partial g^{\lambda\eta}}\right],
\end{eqnarray}
where $f_{\mathcal{R}}=\frac{\partial
f(\mathcal{R},\mathcal{T},\mathcal{Q})}{\partial
\mathcal{R}},~f_{\mathcal{T}}=\frac{\partial
f(\mathcal{R},\mathcal{T},\mathcal{Q})}{\partial \mathcal{T}}$ and
$f_{\mathcal{Q}}=\frac{\partial
f(\mathcal{R},\mathcal{T},\mathcal{Q})}{\partial \mathcal{Q}}$.
Moreover, the symbol $\nabla_\chi$ defines the covariant derivative
and $\Box\equiv g^{\chi\xi}\nabla_\chi\nabla_\xi$. We assume
$\mathcal{L}_{m}=-\mu$ in this case, $\mu$ indicates the energy
density of the fluid which leads to
$\frac{\partial^2\mathcal{L}_{m}} {\partial g^{\chi\xi}\partial
g^{\lambda\eta}}=0$ \cite{23}. Due to the presence of arbitrary
coupling between matter and geometry, the divergence of
$\mathrm{EMT}$ (i.e., $\nabla_\chi \mathcal{T}^{\chi\xi}\neq 0$) in
this theory does not disappear unlike GR and $f(\mathcal{R})$
theory. Thus the equivalence principle is violated due to which
there exists an additional force in the structure which prevents the
moving particles to obey geodesic path in the gravitational field.
Hence we get
\begin{align}\nonumber
\nabla^\chi
\mathcal{T}_{\chi\xi}&=\frac{2}{2f_\mathcal{T}+\mathcal{R}f_\mathcal{Q}+16\pi}\bigg[\nabla_\chi(f_\mathcal{Q}\mathcal{R}^{\lambda\chi}
\mathcal{T}_{\lambda\xi})-\frac{1}{2}
(f_\mathcal{T}g_{\lambda\eta}+f_\mathcal{Q}\mathcal{R}_{\lambda\eta})\nabla_\xi
\mathcal{T}^{\lambda\eta}\\\label{g11}
&+\nabla_\xi(\mathcal{L}_mf_\mathcal{T})-\mathcal{G}_{\chi\xi}\nabla^\chi(f_\mathcal{Q}\mathcal{L}_m)
-\frac{1}{2}\big\{\nabla^{\chi}(\mathcal{R}f_{\mathcal{Q}})+2\nabla^{\chi}f_{\mathcal{T}}\big\}\mathcal{T}_{\chi\xi}\bigg].
\end{align}

The $\mathrm{EMT}$ characterizes the matter configuration in the
astrophysical structures and each of its non-null components
expresses different physical characteristics. The anisotropy induced
by the difference between pressure components in radial and
tangential directions is observed as an important ingredient to
study the formation and evolution of self-gravitating strange
bodies. There is a large number of massive objects in the universe
which are found to be coupled with anisotropic configuration, thus
this factor has convincing consequences in the evolutionary stages
of stellar structures. We consider anisotropic configured stars for
which the $\mathrm{EMT}$ is defined as
\begin{equation}\label{g5}
\mathcal{T}_{\chi\xi}^{(m)}=(\mu+P_\bot) \mathcal{K}_{\chi}
\mathcal{K}_{\xi}+P_\bot
g_{\chi\xi}+\left(P_r-P_\bot\right)\mathcal{W}_\chi\mathcal{W}_\xi,
\end{equation}
where $P_r$ and $P_\bot$ indicate the radial and tangential
pressures, respectively. Also, $~\mathcal{K}_{\chi}$ is the
four-velocity and $\mathcal{W}_\chi$ denotes the four-vector. The
field equations in $f(\mathcal{R},\mathcal{T},\mathcal{Q})$ theory
provide the trace as
\begin{align}\nonumber
&3\nabla^{\lambda}\nabla_{\lambda}
f_\mathcal{R}+\mathcal{R}\left(f_\mathcal{R}-\frac{\mathcal{T}}{2}f_\mathcal{Q}\right)-\mathcal{T}(f_\mathcal{T}+8\pi)+\frac{1}{2}
\nabla^{\lambda}\nabla_{\lambda}(f_\mathcal{Q}\mathcal{T})+\nabla_\chi\nabla_\lambda(f_\mathcal{Q}\mathcal{T}^{\chi\lambda})\\\nonumber
&-2f+(\mathcal{R}f_\mathcal{Q}+4f_\mathcal{T})\mathcal{L}_m+2\mathcal{R}_{\chi\lambda}\mathcal{T}^{\chi\lambda}f_\mathcal{Q}
-2g^{\xi\eta} \frac{\partial^2\mathcal{L}_m}{\partial
g^{\xi\eta}\partial
g^{\chi\lambda}}\left(f_\mathcal{T}g^{\chi\lambda}+f_\mathcal{Q}R^{\chi\lambda}\right)=0.
\end{align}
In a stellar object, the strong matter-geometry coupling disappears
by assuming $\mathcal{Q}=0$ in the overhead equation, thus we get
$f(\mathcal{R},\mathcal{T})$ theory, whereas the consideration of
vacuum scenario provides the $f(\mathcal{R})$ theory.

The spherically symmetric geometry under consideration contains
inner and outer regions separated by the hypersurface $\Sigma$. We
take a metric which expresses static matter configuration
corresponding to the inner spacetime as follows
\begin{equation}\label{g6}
ds^2=-e^{\phi} dt^2+e^{\psi} dr^2+r^2d\theta^2+r^2\sin^2\theta
d\varphi^2,
\end{equation}
where $\phi=\phi(r)$ and $\psi=\psi(r)$. We assume the comoving
framework for our analysis, thus the four-velocity and four-vector
have the only non-zero components as
\begin{equation}\label{g7}
\mathcal{K}^\chi=\delta^\chi_0 e^{\frac{-\phi}{2}}, \quad
\mathcal{W}^\chi=\delta^\chi_1 e^{\frac{-\psi}{2}},
\end{equation}
which must satisfy $\mathcal{K}^\chi \mathcal{K}_{\chi}=-1$ and
$\mathcal{W}^\chi \mathcal{K}_{\chi}=0$. There exist numerous stars
in non-linear regime in the current evolutionary phase of our
universe. We need to study the linear behavior of such objects to
obtain a comprehensive description of their structural formation. As
this theory encompasses the more complex functional, we thus adopt a
separable model suggested by Haghani \emph{et al.} \cite{22} to
analyze the influence of
$\mathcal{Q}=\mathcal{R}_{\chi\xi}\mathcal{T}^{\chi\xi}$ on
different quark candidates as
\begin{equation}\label{g61}
f(\mathcal{R},\mathcal{T},\mathcal{Q})=f_1(\mathcal{R})+
f_2(\mathcal{Q}).
\end{equation}
We consider $f_1(\mathcal{R})=\mathcal{R}$ and
$f_2(\mathcal{Q})=\varrho\mathcal{Q}$, where $\varrho$ is an
arbitrary coupling constant.

It is noticeable that different choices of the coupling parameters
for physically feasible models should lie in their observed limits.
This model has widely been used to study the stability and viability
of various anisotropic solutions \cite{22a,22b,25a}. Here,
\begin{eqnarray}\nonumber
\mathcal{Q}&=&e^{-\psi}\bigg[\frac{\mu}{4}\left(\phi'^2-\phi'\psi'+2\phi''+\frac{4\phi'}{r}\right)-\frac{P_r}{4}\left(\phi'^2-\phi'\psi'
+2\phi''+\frac{4\psi'}{r}\right)\\\nonumber &+&P_\bot
\left(\frac{\psi'}{r}-\frac{\phi'}{r}+\frac{2e^\psi}{r^2}-\frac{2}{r^2}\right)\bigg].
\end{eqnarray}
By inserting a particular model \eqref{g61} in Eq.\eqref{g2} and
combining it with Eq.\eqref{g4}, we obtain
\begin{eqnarray}\nonumber
\mathcal{G}_{\chi\xi}&=&\frac{1}{\varrho\mu+1}
\bigg[\left(8\pi+\frac{1}{2}\varrho\mathcal{R}\right)\mathcal{T}_{\chi\xi}^{(m)}
+\frac{\varrho}{2}\left\{
\mathcal{Q}-\nabla_{\lambda}\nabla_{\eta}\mathcal{T}^{\lambda\eta}\right\}g_{\chi\xi}-\frac{\varrho}{2}\Box\mathcal{T}_{\chi\xi}\\\label{g7a}
&-&2\varrho\mathcal{R}_{\lambda(\chi}\mathcal{T}_{\xi)}^{\lambda}+\varrho\nabla_{\lambda}\nabla_{(\chi}\mathcal{T}_{\xi)}^{\lambda}\bigg].
\end{eqnarray}
The covariant divergence \eqref{g11} for the considered model takes
the form
\begin{align}\label{g7b}
\nabla^\chi
\mathcal{T}_{\chi\xi}=\frac{2\varrho}{\varrho\mathcal{R}+16\pi}\left[\nabla_\chi(\mathcal{R}^{\lambda\chi}\mathcal{T}_{\lambda\xi})-\frac{1}{2}
\mathcal{R}_{\lambda\eta}\nabla_\xi
\mathcal{T}^{\lambda\eta}-\frac{1}{2}
\mathcal{T}_{\chi\xi}\nabla^\chi\mathcal{R}-\mathcal{G}_{\chi\xi}\nabla^\chi\mathcal{L}_m\right].
\end{align}
By utilizing Eqs.\eqref{g5} and \eqref{g7a} along with geometry
\eqref{g6}, the field equations in this theory become
\begin{align}\nonumber
8\pi\mu&=e^{-\psi}\bigg[\frac{\psi'}{r}+\frac{e^\psi}{r^2}-\frac{1}{r^2}+\varrho\bigg\{\mu\bigg(\frac{3\phi'\psi'}{8}-\frac{\phi'^2}{8}
+\frac{2\psi'}{r}+\frac{2e^\psi}{r^2}-\frac{2}{r^2}-\frac{3\phi''}{4}\\\nonumber
&-\frac{3\phi'}{2r}\bigg)-\mu'\bigg(\frac{\psi'}{4}-\frac{1}{r}-\phi'\bigg)+\frac{\mu''}{2}-P_r\bigg(\frac{\phi'^2}{8}
-\frac{\phi'\psi'}{8}+\frac{\phi''}{4}-\frac{\psi'}{2r}-\frac{\psi''}{2}\\\label{g8}
&+\frac{3\psi'^2}{4}\bigg)+\frac{5\psi'P'_r}{4}-\frac{P''_r}{2}+P_\bot\bigg(\frac{\psi'}{2r}-\frac{\phi'}{2r}+\frac{e^\psi}{r^2}
+\frac{1}{r^2}\bigg)-\frac{P'_\bot}{r}\bigg\}\bigg],\\\nonumber 8\pi
P_r&=e^{-\psi}\bigg[\frac{\phi'}{r}-\frac{e^\psi}{r^2}+\frac{1}{r^2}+\varrho\bigg\{\mu\bigg(\frac{\phi'\psi'}{8}+\frac{\phi'^2}{8}
-\frac{\phi''}{4}+\frac{\phi'}{2r}-\frac{e^\psi}{r^2}+\frac{1}{r^2}\bigg)\\\nonumber
&-\frac{\phi'\mu'}{4}-P_r\bigg(\frac{5\phi'^2}{8}-\frac{7\phi'\psi'}{8}+\frac{5\phi''}{4}-\frac{7\psi'}{2r}+\frac{\phi'}{r}-\psi'^2
-\frac{e^\psi}{r^2}+\frac{1}{r^2}\bigg)+P'_r\\\label{g9}
&\times\bigg(\frac{\phi'}{4}+\frac{1}{r}\bigg)-P_\bot\bigg(\frac{\psi'}{2r}-\frac{\phi'}{2r}+\frac{e^\psi}{r^2}
+\frac{1}{r^2}\bigg)+\frac{P'_\bot}{r}\bigg\}\bigg],\\\nonumber 8\pi
P_\bot&=e^{-\psi}\bigg[\frac{\phi'^2}{4}-\frac{\phi'\psi'}{4}+\frac{\phi''}{2}-\frac{\psi'}{2r}+\frac{\phi'}{2r}
+\varrho\bigg\{\mu\bigg(\frac{3\phi'^2}{8}-\frac{\phi'\psi'}{8}
+\frac{\phi''}{4}-\frac{\psi'}{2r}\bigg)\\\nonumber
&-\frac{\phi'\mu'}{4}+P_r\bigg(\frac{\phi'^2}{8}-\frac{\phi'\psi'}{8}+\frac{\phi''}{4}-\frac{\psi'}{2r}-\frac{\psi''}{2}
+\frac{3\psi'^2}{4}\bigg)-\frac{5\psi'P'_r}{4}+\frac{P''_r}{2}\\\nonumber
&-P_\bot\bigg(\frac{\phi'^2}{4}-\frac{\phi'\psi'}{4}+\frac{\phi''}{2}-\frac{\psi'}{r}+\frac{\phi'}{r}-\frac{2e^\psi}{r^2}
+\frac{1}{r^2}\bigg)-P'_\bot\bigg(\frac{\psi'}{4}-\frac{\phi'}{4}-\frac{3}{r}\bigg)\\\label{g10}
&+\frac{P''_\bot}{2}\bigg\}\bigg],
\end{align}
where the matter variables on the right side of above equations
appear due to the modified gravity which make the system more
complicated. Here, prime symbolizes $\frac{\partial}{\partial r}$.
The expression for hydrostatic equilibrium in
$f(\mathcal{R},\mathcal{T},\mathcal{Q})$ scenario is obtained with
the help of Eq.\eqref{g7b} as
\begin{align}\nonumber
&\frac{dP_r}{dr}+\frac{\phi'}{2}\left(\mu
+P_r\right)-\frac{2}{r}\left(P_\bot-P_r\right)-\frac{2\varrho
e^{-\psi}}{\varrho\mathcal{R}+16\pi}\bigg[\frac{\phi'\mu}{8}\bigg(\phi'^2-\phi'\psi'+2\phi''+\frac{4\phi'}{r}\bigg)\\\nonumber
&-\frac{\mu'}{8}\bigg(\phi'^2-\phi'\psi'+2\phi''-\frac{4\phi'}{r}-\frac{8e^\psi}{r^2}+\frac{8}{r^2}\bigg)+P_r\bigg(\frac{5\phi'^2\psi'}{8}
-\frac{5\phi'\psi'^2}{8}-\frac{5\psi'^2}{2r}\\\nonumber
&+\frac{7\phi''\psi'}{4}+\frac{\phi'\psi''}{2}-\phi'\phi''-\frac{\phi'''}{2}+\frac{2\psi''}{r}+\frac{\phi'\psi'}{r}-\frac{\psi'}{r^2}
-\frac{\phi''}{r}+\frac{\phi'}{r^2}+\frac{2e^\psi}{r^3}-\frac{2}{r^3}\bigg)\\\nonumber
&+\frac{P'_r}{8}\bigg(\phi'\psi'-\phi'^2-2\phi''+\frac{4\psi'}{r}\bigg)+\frac{P_\bot}{r^2}\bigg(\psi'-\phi'+\frac{2e^\psi}{r}
-\frac{2}{r}\bigg)-\frac{P'_\bot}{r}\bigg(\frac{\psi'}{2}\\\label{g12}
&-\frac{\phi'}{2}+\frac{e^\psi}{r}-\frac{1}{r}\bigg)\bigg]=0.
\end{align}
The generalization of Tolman-Opphenheimer-Volkoff ($\mathrm{TOV}$)
equation in this theory is illustrated by Eq.\eqref{g12}. This
equation seems to be very significant in interrogating the
structural evolution of self-gravitating bodies.

The matter variables of the fluid distribution can be interlinked
through different constraints, known as equations of state which
help to study the physical aspects of a stellar body. The most
fascinating objects in our universe are the neutron stars which are
formed after the collapse of heavily structures having masses 8 to
20 times mass of the sun. The sufficiently dense stars can further
be turned into black holes, while the less dense transform into
quark stars whose conversion has been examined by various
researchers \cite{33b,41c}. These stars are found to be small in
size, highly dense and occupy strong gravitational field. Due to
non-linearity in the field equations \eqref{g8}-\eqref{g10}
involving five unknowns $(\phi,\psi,\mu,P_r,P_\bot)$, we need some
constraints to make the system solvable. We suppose that the matter
variables in the interior of compact models are interlinked through
MIT bag model EoS which plays a considerable role to analyze quark
stars \cite{27}. We define the quark pressure as
\begin{equation}\label{g13}
P_r=\sum_{k=u,d,s}P^k-\mathfrak{B_c},
\end{equation}
where $\mathfrak{B_c}$ indicates the bag constant. Also, the
pressures $P^u,~P^d$ and $P^s$ correspond to the up, down and
strange quark matters, respectively. Each quark density is
interlinked with respective quark pressure as $\mu^k=3P^k$. Thus the
energy density is expressed as
\begin{equation}\label{g14}
\mu=\sum_{k=u,d,s}\mu^k+\mathfrak{B_c}.
\end{equation}
We construct MIT bag model EoS which illustrates the strange matter
by combining Eqs.\eqref{g13} and \eqref{g14} as
\begin{equation}\label{g14a}
P_r=\frac{1}{3}\left(\mu-4\mathfrak{B_c}\right).
\end{equation}
Various authors \cite{41f,41g} analyzed the physical characteristics
of quark stars successfully by taking different values of the bag
constant for the above EoS. Our main purpose is to find analytic
solution of the field equations, thus after using the EoS
\eqref{g14a} in Eqs.\eqref{g8}-\eqref{g10}, we have
\begin{align}\nonumber
\mu&=\bigg[8\pi
e^{\psi}+\varrho\bigg(\frac{9\phi''}{8}-\frac{e^{\psi}}{r^2}+\frac{1}{r^2}-\frac{\psi''}{8}-\frac{5\phi'\psi'}{8}-\frac{\psi'^2}{16}
-\frac{5\psi'}{2r}+\frac{3\phi'^2}{16}+\frac{\phi'}{r}\bigg)\bigg]^{-1}\\\nonumber
&\times\bigg[\frac{3}{4}\bigg(\frac{\psi'}{r}+\frac{\phi'}{r}\bigg)+\mathfrak{B_c}\bigg\{8\pi
e^\psi-\varrho\bigg(\frac{\psi''}{2}
+\frac{4\psi'}{r}-\frac{3\phi'^2}{4}-\frac{3\phi''}{2}+\frac{\psi'^2}{4}-\frac{\phi'}{r}\\\label{g14b}
&+\frac{e^\psi}{r^2}+\phi'\psi'-\frac{1}{r^2}\bigg)\bigg\}\bigg],\\\nonumber
P_r&=\bigg[8\pi
e^{\psi}+\varrho\bigg(\frac{9\phi''}{8}-\frac{e^{\psi}}{r^2}+\frac{1}{r^2}-\frac{\psi''}{8}-\frac{5\phi'\psi'}{8}-\frac{\psi'^2}{16}
-\frac{5\psi'}{2r}+\frac{3\phi'^2}{16}+\frac{\phi'}{r}\bigg)\bigg]^{-1}\\\label{g14c}
&\times\bigg[\frac{1}{4}\bigg(\frac{\psi'}{r}+\frac{\phi'}{r}\bigg)-\mathfrak{B_c}\bigg\{8\pi
e^\psi-\varrho\bigg(\frac{\phi'\psi'}{2}
+\frac{2\psi'}{r}-\frac{\phi'}{r}+\frac{e^\psi}{r^2}-\phi''-\frac{1}{r^2}\bigg)\bigg\}\bigg],\\\nonumber
P_\bot&=\bigg[8\pi
e^{\psi}+\varrho\bigg(\frac{\phi'^2}{4}-\frac{2e^{\psi}}{r^2}+\frac{1}{r^2}-\frac{\phi'\psi'}{4}+\frac{\phi''}{2}-\frac{\psi'}{r}
+\frac{\phi'}{r}\bigg)\bigg]^{-1}\bigg[\frac{\phi'^2}{4}-\frac{\psi'}{2r}\\\nonumber
&+\frac{\phi'}{2r}-\frac{\phi'\psi'}{4}+\frac{\phi''}{2}+\varrho\bigg\{8\pi
e^{\psi}+\varrho\bigg(\frac{9\phi''}{8}-\frac{e^{\psi}}{r^2}+\frac{1}{r^2}-\frac{\psi''}{8}
-\frac{5\phi'\psi'}{8}-\frac{\psi'^2}{16}\\\nonumber
&-\frac{5\psi'}{2r}+\frac{3\phi'^2}{16}+\frac{\phi'}{r}\bigg)\bigg\}^{-1}\bigg\{\frac{1}{16r}\bigg(\phi'\psi'^2+3\phi'^2\psi'+5\phi'^3+4\phi''\psi'
+4\phi'\phi''\\\nonumber
&-2\psi'\psi''-2\phi'\psi''+3\psi'^3-\frac{8\psi'^2}{r}-\frac{8\phi'\psi'}{r}\bigg)+2\pi
e^\psi\mathfrak{B_c}\bigg(\phi'^2+2\psi'' -3\psi'^2\bigg)\\\nonumber
&+\frac{\varrho\mathfrak{B_c}}{16}\bigg(4\phi''\psi''-11\phi''\psi'^2-3\phi'^2\psi''+10\phi''\phi'^2-3\phi'\psi'\psi''-3\phi'\phi''\psi'\\\nonumber
&+2\phi''^2+\frac{5\phi'^2\psi'^2}{2}-\frac{13\phi'^3\psi'}{2}-\frac{4\phi'\psi'^2}{r}
+\frac{11\phi'\psi'^3}{2}-\frac{26\phi'^2\psi'}{r}-\frac{12\phi''\psi'}{r}\\\nonumber
&+\frac{4\phi'^3}{r}-\frac{12\psi'\psi''}{r}+\frac{22\psi'^3}{r}+\frac{9\phi'^4}{2}-\frac{4\phi'^2e^\psi}{r^2}+\frac{4\phi'^2}{r^2}-\frac{8\psi''e^\psi}{r^2}
-\frac{12\psi'^2e^\psi}{r^2}\\\label{g14d}
&+\frac{8\psi''}{r^2}+\frac{4\psi'^2}{r^2}\bigg)\bigg\}\bigg].
\end{align}

\subsection{Krori-Barua Solution}

It is noticed that various researchers utilized the EoS \eqref{g14a}
to explore the physical features of different quark stars in both GR
as well as modified framework. Our aim is to develop anisotropic
solution by means of such a simplest EoS and analyze its feasibility
corresponding to five star candidates. To do this, we consider
Krori-Barua solution \cite{41h} in
$f(\mathcal{R},\mathcal{T},\mathcal{Q})$ scenario which has acquired
a lot of attention due to its singularity free nature. The solution
has the form
\begin{equation}\label{g15}
\phi=Br^2+C, \quad \psi=Ar^2,
\end{equation}
where $A,~B$ and $C$ are unknowns and their values can be calculated
through matching conditions. Now, we check the criteria for
acceptability of these metric coefficients \cite{41j}, thus there
derivatives upto second order are
\begin{align}\nonumber
\phi'(r)=2Br,\quad \psi'(r)=2Ar,\quad \phi''(r)=2B,\quad
\psi''(r)=2A,
\end{align}
from where we observe that $\phi'(0)=\psi'(0)=0,~\phi''(0)>0$ and
$\psi''(0)>0$ everywhere ($r=0$ is center of the star). Hence both
the metric potentials given in Eq.\eqref{g15} are acceptable. The
field equations \eqref{g14b}-\eqref{g14d} in the Krori-Barua
framework \eqref{g15} become
\begin{align}\nonumber
\mu&=\bigg[\varrho\left(-A^2r^4-Ar^2\left(10Br^2+21\right)+3B^2r^4+17Br^2+4\right)+32\pi
r^2e^{Ar^2}\\\nonumber &-4\varrho
e^{Ar^2}\bigg]^{-1}\bigg[2\left(-2\varrho A^2\mathfrak{B_c}
r^4+Ar^2\left(3-2\varrho\mathfrak{B_c}\left(4Br^2+9\right)\right)+2\mathfrak{B_c}\left(\varrho\right.\right.\\\label{g16}
&+\left.\left.e^{Ar^2}\left(8\pi
r^2-\varrho\right)\right)+6\varrho\mathfrak{B_c}B^2r^4+Br^2(10\varrho\mathfrak{B_c}+3)\right)\bigg],\\\nonumber
P_r&=\bigg[\varrho\left(-A^2r^4-Ar^2\left(10Br^2+21\right)+3B^2r^4+17Br^2+4\right)+32\pi
r^2e^{Ar^2}\\\nonumber &-4\varrho
e^{Ar^2}\bigg]^{-1}\bigg[2\left(Ar^2\left(4\varrho\mathfrak{B_c}\left(Br^2+2\right)+1\right)-2\mathfrak{B_c}\left(\varrho+e^{Ar^2}\left(8\pi
r^2-\varrho\right)\right)\right.\\\label{g17}
&\left.+Br^2(1-8\varrho\mathfrak{B_c})\right)\bigg],\\\nonumber
P_\bot&=\bigg[\bigg\{\varrho\left(-Ar^2\left(Br^2+2\right)+B^2r^4+3Br^2+1\right)+e^{Ar^2}
\left(8\pi
r^2-\varrho\right)\bigg\}\bigg\{\varrho\left(4\right.\\\nonumber
&-\left.A^2r^4-Ar^2\left(10Br^2+21\right)+3B^2r^4+17Br^2\right)+4e^{A
r^2}\left(8\pi r^2-\varrho\right)\bigg\}\bigg]^{-1}\\\nonumber
&\times\bigg[r^2\bigg\{\varrho
A^3r^4\left(44\varrho\mathfrak{B_c}+Br^2(22\varrho\mathfrak{B_c}+1)+7\right)-\varrho
A^2r^2\left(8\varrho\mathfrak{B_c}+4\mathfrak{B_c}
e^{Ar^2}\right.\\\nonumber &\times\left.\left(\varrho+24\pi
r^2\right)-B^2r^4 (10\varrho\mathfrak{B_c}+9)-Br^2
(31-36\varrho\mathfrak{B_c})-11\right)+B\left(4e^{A
r^2}\right.\\\nonumber &\times\left(8\pi
r^2-\varrho\right)\left(Br^2(\varrho\mathfrak{B_c}+1)+2\right)+\varrho
\left(3B^3r^6(6\varrho\mathfrak{B_c}+1)+B^2r^4
(28\varrho\mathfrak{B_c}\right.\\\nonumber
&+\left.\left.33)+6Br^2(\varrho\mathfrak{B_c}+7)+8\right)\right)-A\left(4e^{A
r^2}\left(8\pi
r^2-\varrho\right)\left(-\varrho\mathfrak{B_c}+Br^2+1\right)\right.\\\nonumber
&+\varrho\left(13B^3r^6(2\varrho\mathfrak{B_c}+1)+B^2r^4(64\varrho\mathfrak{B_c}+55)+Br^2(8\varrho\mathfrak{B_c}+69)
-4\varrho\mathfrak{B_c}\right.\\\label{g18}
&+\left.\left.4\right)\right)\bigg\}\bigg].
\end{align}

\section{Boundary Conditions}

To analyze the exact structural configuration of anisotropic compact
stars, the existence of smooth matching between inner and outer
geometries plays significant role. We take outer Schwarzschild
spacetime in this context which is symbolized by the metric as
\begin{equation}\label{g20}
ds^2=-\left(1-\frac{2\bar{M}}{r}\right)dt^2+\left(1-\frac{2\bar{M}}{r}\right)^{-1}dr^2+r^2d\theta^2+r^2\sin^2\theta
d\varphi^2,
\end{equation}
where $\bar{M}(r)$ indicates the total mass of a star at the
boundary ($r=\mathcal{H}$). The continuity of the metric
coefficients of both geometries at boundary surface produces some
constraints as
\begin{eqnarray}\label{g21}
g_{tt}&=&e^{B\mathcal{H}^2+C}=1-\frac{2\bar{M}}{\mathcal{H}}, \quad
g_{rr}=e^{-A\mathcal{H}^2}=1-\frac{2\bar{M}}{\mathcal{H}},\\\label{g22}
\frac{\partial g_{tt}}{\partial
r}&=&B\mathcal{H}e^{B\mathcal{H}^2+C}=\frac{\bar{M}}{\mathcal{H}^2}.
\end{eqnarray}
After solving the above three equations simultaneously, we obtain
the values of triplet ($A,B,C$) as
\begin{eqnarray}\label{g23}
A&=&-\frac{1}{\mathcal{H}^2}
\ln\left(1-\frac{2\bar{M}}{\mathcal{H}}\right),\\\label{g24}
B&=&\frac{\bar{M}}{\mathcal{H}^3}
\left(1-\frac{2\bar{M}}{\mathcal{H}}\right)^{-1},\\\label{g25}
C&=&\ln\left(1-\frac{2\bar{M}}{\mathcal{H}}\right)
-\frac{\bar{M}}{\mathcal{H}-2\bar{M}}.
\end{eqnarray}
The radial pressure in stellar structures must vanish at the
boundary ($r=\mathcal{H}$), thus Eq.\eqref{g17} along with
Eqs.\eqref{g23}-\eqref{g25} lead to the following expression
\begin{align}\nonumber
P_r|_{(r=\mathcal{H})}&=\bigg[15\varrho
\bar{M}^2+\varrho\left(-64\bar{M}^2+74\bar{M}\mathcal{H}-21
\mathcal{H}^2\right)\log\left(1-\frac{2\bar{M}}{\mathcal{H}}\right)+64\pi\bar{M}\mathcal{H}^3\\\nonumber
&-9\varrho\bar{M}\mathcal{H}+\varrho(\mathcal{H}-2\bar{M})^2\log^2\left(1-\frac{2\bar{M}}{\mathcal{H}}\right)-32\pi
\mathcal{H}^4\bigg]^{-1}\bigg[2(2\bar{M}-\mathcal{H})\\\nonumber
&\times\big\{-4\varrho\mathfrak{B_c}\bar{M}+\log\left(1-\frac{2\bar{M}}{\mathcal{H}}\right)(2\bar{M}(6\varrho\mathfrak{B_c}+1)
-\mathcal{H}(8\varrho\mathfrak{B_c}+1))\\\label{g26}
&+\bar{M}-16\pi\mathfrak{B_c}\mathcal{H}^3\big\}\bigg]=0.
\end{align}
The bag constant can be evaluated from Eq.\eqref{g26} as
\begin{align}\label{g27}
\mathfrak{B_c}=\frac{(2\bar{M}-\mathcal{H})\log\left(1-\frac{2\bar{M}}{\mathcal{H}}\right)+\bar{M}}{2
\left\{2\varrho\bar{M}+2\varrho(2\mathcal{H}-3\bar{M})\log\left(1-\frac{2\bar{M}}{\mathcal{H}}\right)+8\pi\mathcal{H}^3\right\}}.
\end{align}
By utilizing the experimental data of different quark stars
\cite{42,42aa}, the values of $A,~B,~C$ and $\mathfrak{B_c}$ can be
determined. These strange bodies are found to be consistent with the
limit proposed by Buchdhal \cite{42a}, i.e.,
$\frac{2\bar{M}}{\mathcal{H}}<\frac{8}{9}$. We choose $\varrho=3$ to
find the value of bag constant for the considered model. This value
of coupling constant helps us in the successful analysis of stellar
evolution. The values of bag constant as well as three unknowns
involved in the Krori-Barua solution corresponding to the observed
masses and radii of considered strange stars are calculated in
Tables $\mathbf{1}$ and $\mathbf{2}$, respectively.

Remarkably, we determine the values of $\mathfrak{B_c}$ for the
different quark stars which are $105.03,~55.28,~200.55,~229.46$ and
$111.73$ $MeV/fm^3$, respectively. The observed values of bag
constant for these stars to be stable are much lesser than the above
calculated values. However, the experimental findings released by
$\mathrm{CERN-SPS}$ and $\mathrm{RHIC}$ present that the density
dependent bag model may yield a vast range of the values of bag
constant.

\section{Physical Analysis of Various Compact Stars}

This section examines different physical features of the considered
strange stars which are coupled with anisotropic configuration in
$f(\mathcal{R},\mathcal{T},\mathcal{Q})$ scenario. We observe the
graphical behavior of matter variables by using the masses and radii
of each star candidate as shown in Table $\mathbf{1}$. We analyze
the viability of metric potentials, energy density, radial and
tangential components of pressure, anisotropy, energy bounds,
compactness as well as redshift for the considered quark candidates
and also investigate their stability, where the model parameter has
been kept fixed. We are familiar with the fact that the
compatibility of a solution guarantees the non-singular and
monotonically increasing nature of metric components, having
positive values throughout. Equation \eqref{g15} shows that the
metric coefficients depend only on Krori-Barua constants. By
utilizing these constants for particular stars calculated in Table
$\mathbf{2}$, the graphical behavior of both metric functions is
analyzed in Figure $\mathbf{1}$ which assures the physical
consistency of the developed solution. It should be noted that the
yellow color expresses 4U 1820-30 compact star, blue indicates Vela
X-I, black represents SAX J 1808.4-3658, green signifies RXJ 1856-37
and red color shows Her X-I in all plots.
\begin{table}[H]
\scriptsize \centering \caption{Physical values of different compact
star candidates} \label{Table1} \vspace{+0.1in}
\setlength{\tabcolsep}{0.95em}
\begin{tabular}{cccccc}
\hline\hline Star Models & 4U 1820-30 & Vela X-I & SAX J 1808.4-3658
& RXJ 1856-37 & Her X-I
\\\hline $Mass(M_{\bigodot})$ & 2.25 & 1.77 & 1.435 & 0.9041 & 0.88
\\\hline
$R(km)$ & 10 & 12.08 & 7.07 & 6 & 7.7
\\\hline
$\frac{M}{R}$ & 0.331 & 0.215 & 0.298 & 0.222 & 0.168
\\\hline
$\mathfrak{B_c}$ & 0.000139001 & 0.000073158 & 0.000265408 & 0.000303665 & 0.000147867  \\
\hline\hline
\end{tabular}
\end{table}
\begin{table}[H]
\scriptsize \centering \caption{Calculated values of Krori-Barua
constants $A,~B$ and $C$ for different compact star candidates}
\label{Table2} \vspace{+0.1in} \setlength{\tabcolsep}{0.95em}
\begin{tabular}{cccccc}
\hline\hline Star Models & 4U 1820-30 & Vela X-I & SAX J 1808.4-3658
& RXJ 1856-37 & Her X-I
\\\hline $A$ & 0.0108323 & 0.0038614 & 0.0181686 & 0.0162557 & 0.0069063
\\\hline
$B$ & 0.0097711 & 0.0025930 & 0.0148019 & 0.0110467 & 0.0042674
\\\hline
$C$ & -2.06034 & -0.94188 & -1.64803 & -0.98289 & -0.66249 \\
\hline\hline
\end{tabular}
\end{table}

\subsection{Inspection of Physical Parameters}

The composition of compact gravitational bodies indicates that the
energy density and both pressure ingredients should be maximum
inside the stellar structure. Figure $\mathbf{2}$ exhibits the
variation in these parameters with respect to each quark candidate
for the model \eqref{g61}. These graphs clearly demonstrate that the
energy density and pressure components gain their maximum values at
the center ($r=0$) of anisotropic configured stars, resulting in the
existence of extremely dense structures. Figure $\mathbf{2}$
\textbf{(b)} also shows that the radial pressure inside the
considered strange stars disappear at the boundary, while the energy
density and tangential pressure decrease linearly with the rise in
$r$. The matter variables fulfill $\frac{d\mu}{dr} <
0,~\frac{dP_r}{dr} < 0$ and $\frac{dP_\bot}{dr} < 0$ as shown in
Figure $\mathbf{3}$, thus they yield regular behavior. As a result,
we observe from this graphical analysis that there must exist highly
compact stars having anisotropic configuration in
$f(\mathcal{R},\mathcal{T},\mathcal{Q})$ gravity.
\begin{figure}\center
\begin{tabular}{ccc}
\includegraphics[width=0.47\textwidth]{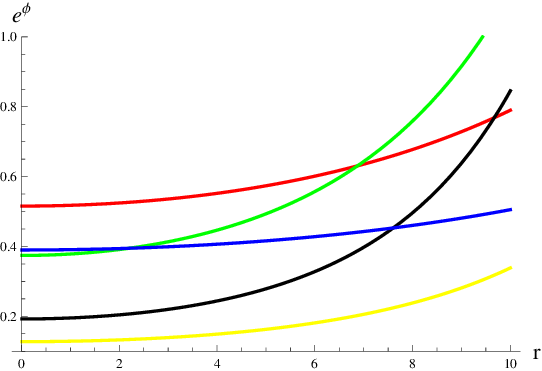} & \includegraphics[width=0.47\textwidth]{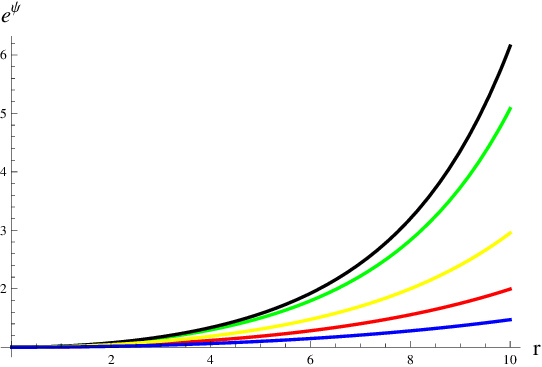}\\
\textbf{(a)} & \textbf{(b)}
\end{tabular}
\caption{Behavior of temporal \textbf{(a)} and radial \textbf{(b)}
metric potentials versus $r$ for different compact star candidates}
\end{figure}

\subsection{Effect of Anisotropic Pressure}

We calculate the anisotropic factor in terms of Krori-Barua ansatz
and bag constant $\mathfrak{B_c}$ by making use of Eqs.\eqref{g17}
and \eqref{g18} as
\begin{align}\nonumber
\Delta&=\bigg[\bigg\{\varrho\left(-Ar^2\left(Br^2+2\right)+B^2r^4+3Br^2+1\right)+e^{Ar^2}
\left(8\pi
r^2-\varrho\right)\bigg\}\bigg\{\varrho\left(4\right.\\\nonumber
&-\left.A^2r^4-Ar^2\left(10Br^2+21\right)+3B^2r^4+17Br^2\right)+4e^{A
r^2}\left(8\pi r^2-\varrho\right)\bigg\}\bigg]^{-1}\\\nonumber
&\times\bigg[\varrho
A^3r^6\left(44\varrho\mathfrak{B_c}+Br^2(22\varrho\mathfrak{B_c}+1)+7\right)+\varrho
A^2r^4\big(3(8\varrho\mathfrak{B_c}+5)-4\mathfrak{B_c}
e^{Ar^2}\\\nonumber &\times\left(\varrho+24\pi
r^2\right)+9B^2r^4(2\varrho\mathfrak{B_c}+1)+Br^2(33-4
\varrho\mathfrak{B_c})\big)+2B^2r^4\big(\varrho(29\varrho\mathfrak{B_c}\\\nonumber
&+18)+2(2\varrho\mathfrak{B_c}+1)e^{Ar^2}\left(8\pi
r^2-\varrho\right)\big)-Ar^2\big(2e^{Ar^2} \left(8\pi
r^2-\varrho\right)\big(3+2Br^2\\\nonumber
&\times(3\varrho\mathfrak{B_c}+1)+10\varrho\mathfrak{B_c}\big)+\varrho\big(20\varrho\mathfrak{B_c}+B^3r^6(34\varrho\mathfrak{B_c}+13)+5
B^2r^4(24\varrho\mathfrak{B_c}\\\nonumber
&+11)+Br^2(71+100\varrho\mathfrak{B_c})+6\big)\big)+4\mathfrak{B_c}\left(\varrho+e^{Ar^2}
\left(8\pi r^2-\varrho\right)\right)^2+2Br^2\\\nonumber
&\times\left(\varrho+e^{A r^2} \left(8\pi
r^2-\varrho\right)\right)(14\varrho\mathfrak{B_c}+3)+3\varrho
B^4r^8(6\varrho\mathfrak{B_c}+1)+\varrho B^3r^6\\\label{g19}
&(44\varrho\mathfrak{B_c}+31)\bigg].
\end{align}
\begin{figure}\center
\begin{tabular}{ccc}
\includegraphics[width=0.47\textwidth]{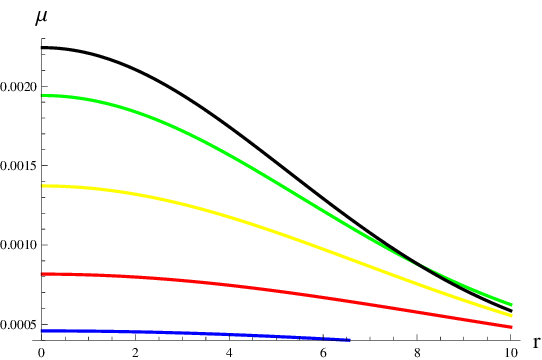} & \includegraphics[width=0.47\textwidth]{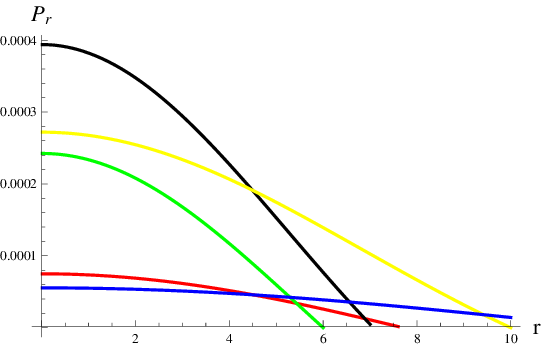}\\
\textbf{(a)} & \textbf{(b)}
\end{tabular}
\begin{tabular}{ccc}
\includegraphics[width=0.47\textwidth]{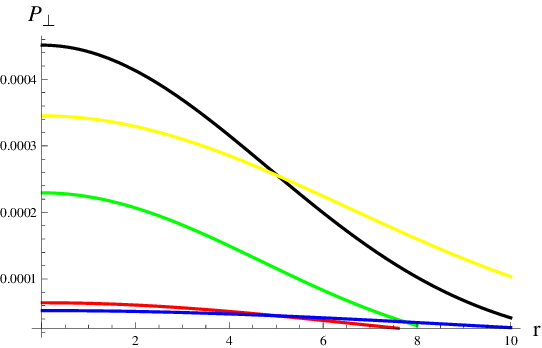}\\
\textbf{(c)}
\end{tabular}
\caption{Plots of energy density (in km$^{-2}$) \textbf{(a)}, radial
pressure (in km$^{-2}$) \textbf{(b)} and tangential pressure (in
km$^{-2}$) \textbf{(c)} versus $r$ for different compact star
candidates}
\end{figure}
We use the observational data of various considered stars (shown in
Table $\mathbf{1}$) to analyze the behavior of anisotropy in their
structural evolution. There occur an outward directed anisotropic
pressure for the case when $P_\bot>P_r$ which yields $\Delta>0$. On
the other hand, the condition $P_\bot<P_r$ (i.e., $\Delta<0$) leads
to the inward directed pressure. The effect of anisotropy on
different stars is shown in Figure $\mathbf{4}$ corresponding to the
viable model of $f(\mathcal{R},\mathcal{T},\mathcal{Q})$ theory. It
is noticed that $\Delta$ remains positive throughout only for 4U
1820-30 and SAX J 1808.4-3658 stars which assures that there exists
a repelling force which contributes to structural evolution of
massive geometries, while this factor varies from negative to
positive in the interior of remaining three candidates.
\begin{figure}\center
\begin{tabular}{ccc}
\includegraphics[width=0.47\textwidth]{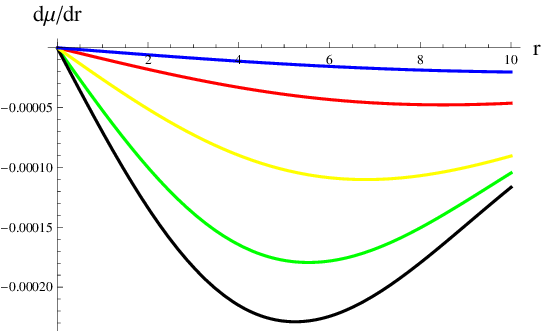} & \includegraphics[width=0.47\textwidth]{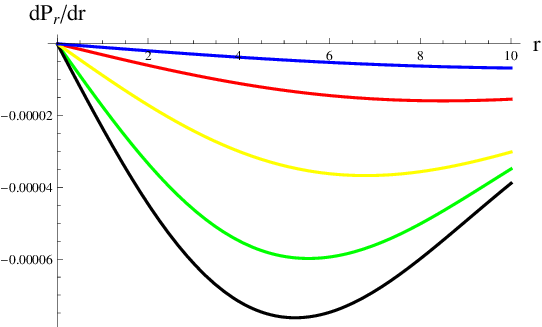}\\
\textbf{(a)} & \textbf{(b)}
\end{tabular}
\begin{tabular}{ccc}
\includegraphics[width=0.47\textwidth]{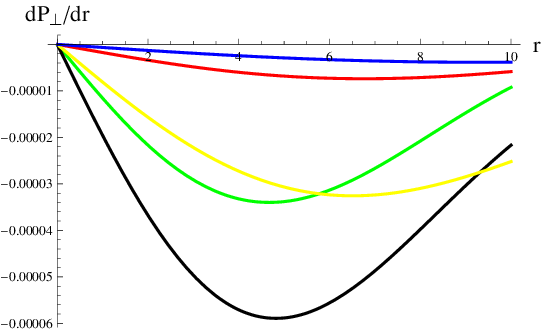}\\
\textbf{(c)}
\end{tabular}
\caption{Plots of $\frac{d\mu}{dr}$ (in km$^{-2}$) \textbf{(a)},
$\frac{dP_r}{dr}$ (in km$^{-2}$) \textbf{(b)} and
$\frac{dP_\bot}{dr}$ (in km$^{-2}$) \textbf{(c)} versus $r$ for
different compact star candidates}
\end{figure}
\begin{figure}\center
\begin{tabular}{ccc}
\includegraphics[width=0.47\textwidth]{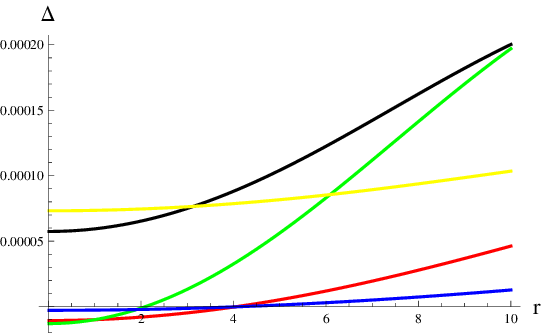} & \includegraphics[width=0.47\textwidth]{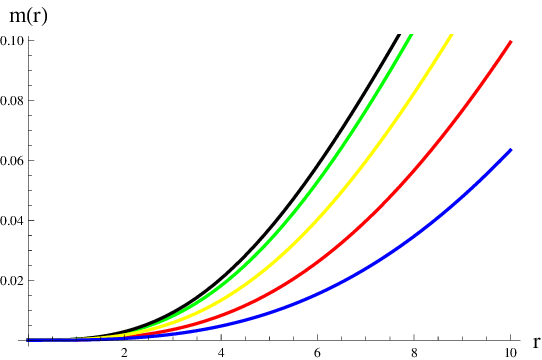}\\
\textbf{(a)} & \textbf{(b)}
\end{tabular}
\caption{Variation of anisotropy (in km$^{-2}$) \textbf{(a)} and
mass (in km) \textbf{(b)} versus $r$ for different compact star
candidates}
\end{figure}

\subsection{Mass, Compactness and Surface Redshift}

For spherical structures, the mass can be defined as
\begin{equation}\label{g63}
m(r)=\frac{1}{2}\int_{0}^{\mathcal{H}}r^2\mu dr,
\end{equation}
where $\mu$ is given in Eq.\eqref{g16}. For our proposed model, we
analyze the graphical behavior of the mass of considered stars by
solving the above equation numerically along with an initial
condition $m(0)=0$, as can be seen from Figure $\mathbf{4}$. We can
characterize a celestial structure by its different physical
features, among them one is the compactness $\big(\sigma(r)\big)$
which defines as the ratio of mass and radius. After employing the
matching conditions between inner and outer spacetimes at
$r=\mathcal{H}$, Buchdahl \cite{42a} found upper bound of
$\sigma(r)$. He disclosed that the system will remain stable for its
value not to be greater than $\frac{4}{9}$. There occur some
reactions in the core of a massive body (having a strong
gravitational force) due to which the electromagnetic radiations
diffuse from that body. The redshift factor measures the increment
in wavelength of those radiations. Mathematically, it is
characterized as
\begin{equation}
z(r)=\frac{1}{\sqrt{1-2\sigma}}-1.
\end{equation}
This factor plays an influential role to study the particles
existing in the inner geometry and its EoS. For perfect fluid
distribution, Buchdahl found its value as $z(r)<2$, whereas Ivanov
\cite{42b} studied anisotropic compact stars and observed its upper
limit to be 5.211. Figure $\mathbf{5}$ shows the plots for
compactness as well as redshift for all quark candidates. It can be
seen that the values of both factors are in their desired ranges.
\begin{figure}\center
\begin{tabular}{ccc}
\includegraphics[width=0.47\textwidth]{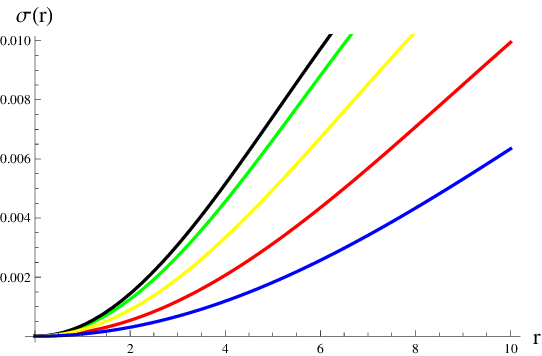} & \includegraphics[width=0.47\textwidth]{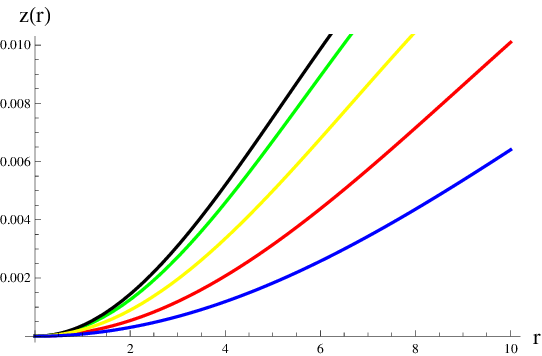}\\
\textbf{(a)} & \textbf{(b)}
\end{tabular}
\caption{Variation of compactness \textbf{(a)} and redshift
\textbf{(b)} factors versus $r$ for different compact star
candidates}
\end{figure}

\subsection{Energy Conditions}

The existence of matter configuration in a stellar body can be
demonstrated by some bounds, known as energy conditions which are of
great importance in astrophysics. We can distinguish the usual or
exotic matter existing inside the geometry by employing such
conditions. They also help to investigate viability of the developed
solutions in any gravitational theory. The physical parameters
representing a particular geometry having ordinary matter must
fulfill these conditions. The energy bounds for anisotropic
configured star in $f(\mathcal{R},\mathcal{T},\mathcal{Q})$ gravity
are
\begin{eqnarray}\nonumber
&&\mu \geq 0, \quad \mu+P_{r} \geq 0,\\\nonumber &&\mu+P_{\bot} \geq
0, \quad \mu-P_{r} \geq 0,\\\label{g50} &&\mu-P_{\bot} \geq 0, \quad
\mu+P_{r}+2P_{\bot} \geq 0.
\end{eqnarray}
The plots of all the above conditions are shown in Figure
$\mathbf{6}$. It is found that these conditions possess positive
trend which assure the viability of the chosen model and the
resulting solution. Thus there must exist normal matter in the
interior of all quark candidates.
\begin{figure}\center
\begin{tabular}{cccc}
\includegraphics[width=0.47\textwidth]{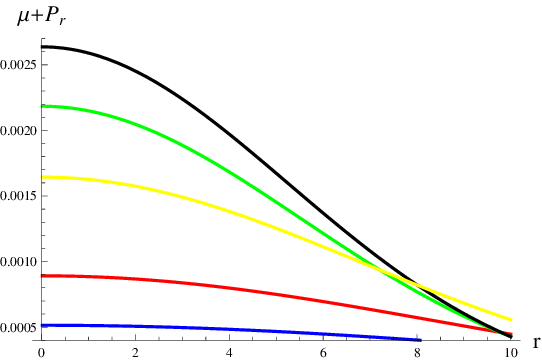} & \includegraphics[width=0.47\textwidth]{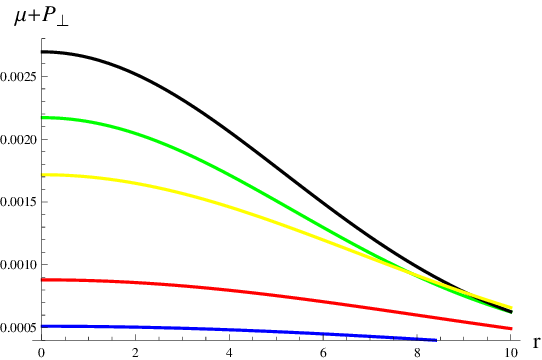}\\
\textbf{(a)} & \textbf{(b)}\\
\includegraphics[width=0.47\textwidth]{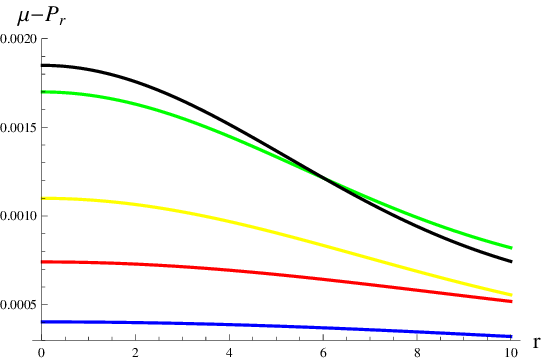} & \includegraphics[width=0.47\textwidth]{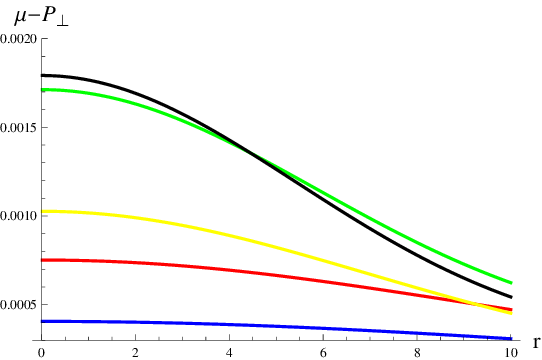}\\
\textbf{(c)} & \textbf{(d)}
\end{tabular}
\begin{tabular}{ccc}
\includegraphics[width=0.47\textwidth]{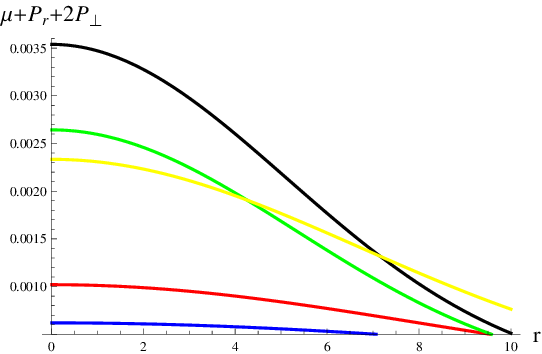}\\
\textbf{(e)}
\end{tabular}
\caption{Plots of energy conditions (in km$^{-2}$) versus $r$ for
different compact star candidates \textbf{(a}$-$\textbf{e)}}
\end{figure}

\subsection{Stability Analysis}

The stability of a compact star attains great significance in
astrophysics to analyze physically feasible models. The massive
bodies which show stable behavior against all the external
fluctuations are more intriguing, thus this phenomenon has
considerable interest in the study of their structural development.
To analyze the stability of considered candidates in
$f(\mathcal{R},\mathcal{T},\mathcal{Q})$ gravity, we employ two
approaches, one of them is the cracking concept presented by Herrera
\cite{29} which is based on sound speed. The causality condition
declares that the squared sound speed $v_{s}^{2}=\frac{dP}{d\mu}$
should lie within $[0,1]$, i.e., $0 \leq v_{s}^{2} < 1$ throughout
for stable structure. This becomes in the case of anisotropic matter
as $0 \leq v_{sr}^{2} < 1$ and $0 \leq v_{s\bot}^{2} < 1$, where
$v_{sr}^{2}=\frac{dP_{r}}{d\mu}$ represents the radial and
$v_{s\bot}^{2}=\frac{dP_{\bot}}{d\mu}$ shows tangential ingredients
of sound speed. Thus the fulfilment of the inequality $0 \leq \mid
v_{s\bot}^{2}-v_{sr}^{2} \mid < 1$ guarantees the stability of
compact object. Figure $\mathbf{7}$ indicates that all the
considered candidates are stable for their respective calculated
values of bag constant and $\varrho=3$.
\begin{figure}\center
\begin{tabular}{ccc}
\includegraphics[width=0.47\textwidth]{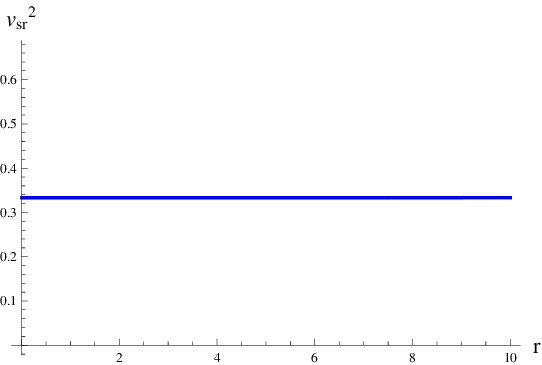} & \includegraphics[width=0.47\textwidth]{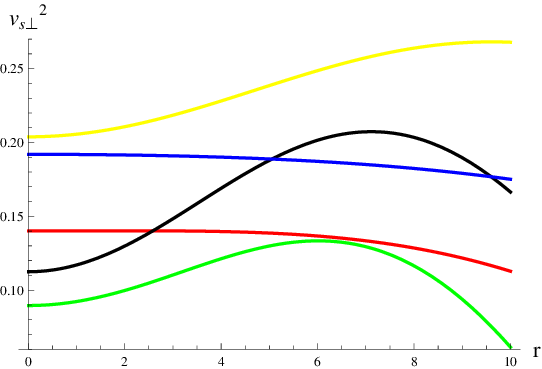}\\
\textbf{(a)} & \textbf{(b)}
\end{tabular}
\begin{tabular}{ccc}
\includegraphics[width=0.47\textwidth]{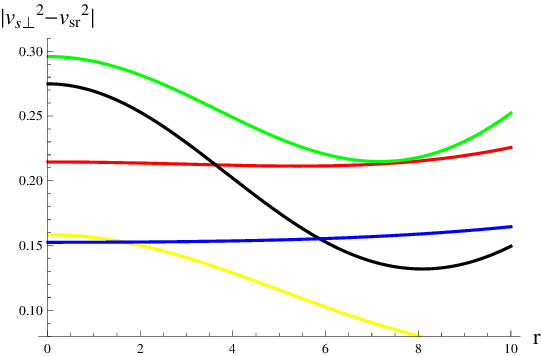}\\
\textbf{(c)}
\end{tabular}
\caption{Plots of $v_{sr}^{2}$ \textbf{(a)}, $v_{s\bot}^{2}$
\textbf{(b)} and $\mid v_{s\bot}^{2}-v_{sr}^{2}\mid$ \textbf{(c)}
versus $r$ for different compact star candidates}
\end{figure}
\begin{figure}\center
\epsfig{file=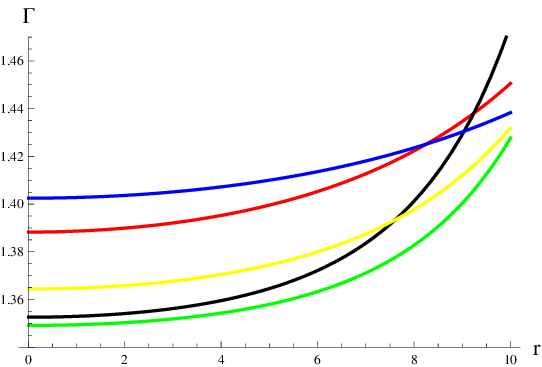,width=0.5\linewidth} \caption{Role of
adiabatic index versus $r$ for different compact star candidates}
\end{figure}

Secondly, the adiabatic index $\big(\Gamma\big)$ is considered as a
powerful tool to analyze the stability of stellar geometry. This
technique has been utilized to study the stable self-gravitating
objects in which the adiabatic index should have its value greater
than $\frac{4}{3}$ everywhere \cite{42d}. In this case, $\Gamma$ is
characterized as
\begin{equation}\label{g62}
\Gamma=\frac{\mu+P_{r}}{P_{r}}
\left(\frac{dP_{r}}{d\mu}\right)=\frac{\mu+P_{r}}{P_{r}}
\left(v_{sr}^{2}\right).
\end{equation}
Figure $\mathbf{8}$ demonstrates the physical behavior of $\Gamma$
for all candidates which fully agrees with the desired limit
throughout the structure.

\section{Conclusions}

This paper discusses the effect of MIT bag constant
($\mathfrak{B_c}$) on physical attributes of five different strange
anisotropic stars, namely 4U 1820-30,~Vela X-I,~SAX J
1808.4-3658,~RXJ 1856-37 and Her X-I in
$f(\mathcal{R},\mathcal{T},\mathcal{Q})$ theory of gravity. We
analyze the influence of strong non-minimal coupling between matter
and geometry in this theory (which appears due to the factor
$\mathcal{Q}=\mathcal{R}_{\chi\xi}\mathcal{T}^{\chi\xi}$) by
adopting a linear model $\mathcal{R}+\varrho \mathcal{Q}$, where the
coupling constant has been kept fixed as $\varrho=3$. We have
formulated the field equations as well as TOV equation with the use
of bag model EoS \eqref{g14a} and also calculated the values of
$\mathfrak{B_c}$ corresponding to each star candidate (Table
$\mathbf{1}$). We have utilized the values of the metric potentials
proposed by Krori-Barua involving three unknowns $(A,B,C)$ whose
values have been evaluated in terms of masses and radii through
matching conditions in this theory. The observational data of
various star candidates has been used to calculate this triplet
(Table $\mathbf{2}$). We have found analytic solution of the
modified field equations by taking two different equations of state
relating energy density with pressure components. The graphical
analysis of all quark candidates has been presented. It is found
that the physical parameters attained their maximum (positive)
values at the center ($r=0$), while the behavior of anisotropy is
increasing towards boundary.

The observed values of redshift and compactness are within their
respective bounds. The energy conditions are fulfilled which confirm
the existence of usual matter inside all the quark candidates as
well as the viability of our developed solution. Two different
techniques have been used to analyze the stability. We have
determined that the potentially stable structure of these stars
exist as the inequalities $0 \leq v_{sr}^{2} < 1,~0 \leq
v_{s\bot}^{2} < 1$ and $0 \leq \mid v_{s\bot}^{2}-v_{sr}^{2} \mid <
1$ hold throughout the system. The adiabatic index for all the
considered stars has been visualized which also assures their stable
structures. It is worthwhile to note that the quark star 4U 1820-30
shows more stable behavior towards the boundary in comparison with
other four candidates (Figure $\mathbf{7}$). It is concluded that
the non-minimal matter-geometry interaction in
$f(\mathcal{R},\mathcal{T},\mathcal{Q})$ theory may yield more
appropriate results for compact structures as compared to
\cite{25a,25b}. We have found that the chosen model \eqref{g61} has
viable behavior as the compact structures obtained with the help of
MIT EoS \eqref{g14a} meet the needed requirements. Finally, we can
retrieve all these results in GR for $\varrho=0$ in
$f(\mathcal{R},\mathcal{T},\mathcal{Q})$ functional form
\eqref{g61}.

\vspace{0.25cm}

\section*{Appendix A}

The value of adiabatic index in terms of Krori-Barua solution takes
the form
\begin{align}\nonumber
\Gamma&=-\bigg[3\left\{Ar^2\left(4\varrho\mathfrak{B_c}\left(Br^2+2\right)+1\right)
-2\mathfrak{B_c}\left(\varrho+e^{Ar^2}\left(8\pi
r^2-\varrho\right)\right)+Br^2\right.\\\nonumber
&\times\left.(1-8\varrho\mathfrak{B_c})\right\}\bigg]^{-1}\bigg[2r^2\left\{\varrho
A^2\mathfrak{B_c}
r^2+A\left(\varrho\mathfrak{B_c}\left(2Br^2+5\right)-2\right)-B\left(2+\varrho\mathfrak{B_c}\right.\right.\\\nonumber
&\times\left.\left.\left(1+3Br^2\right)\right)\right\}\bigg].
\end{align}
The term $\big|v_{s\bot}^2-v_{sr}^2\big|$ in modified gravity
becomes
\begin{align}\nonumber
\big|v_{s\bot}^2-v_{sr}^2\big|&=\frac{1}{12}\bigg|\bigg[\big(\varrho\big(-Ar^2\big(Br^2+2\big)+B^2r^4+3Br^2+1\big)+e^{Ar^2}\big(r^2-\varrho\big)\big)^2\\\nonumber
&\times\big\{\varrho
A^3r^4\big(8\varrho\mathfrak{B_c}+2\mathfrak{B_c}e^{Ar^2}\big(r^2-\varrho\big)+1\big)+A^2r^2\big(e^{Ar^2}\big(2\varrho(2-3\varrho\mathfrak{B_c})\\\nonumber
&+4\varrho\mathfrak{B_c}Br^4-4r^2\big(-2\varrho\mathfrak{B_c}+\varrho^2\mathfrak{B_c}B+1\big)\big)+\varrho\big(Br^2(11-12\varrho\mathfrak{B_c})\\\nonumber
&-4\varrho\mathfrak{B_c}\big)\big)+A\big(\varrho\big(-B^2\big)r^4\big(36\varrho\mathfrak{B_c}+6\mathfrak{B_c}e^{Ar^2}
\big(r^2-\varrho\big)-7\big)-2Br^2\\\nonumber
&\times\big(4\varrho^2\mathfrak{B_c}+e^{Ar^2}\big(r^2(3\varrho\mathfrak{B_c}+2)-\varrho(5\varrho
\mathfrak{B_c}+2)\big)\big)+(5\varrho\mathfrak{B_c}-2)\big(e^{Ar^2}\\\nonumber
&-1\big)2\varrho\big)+\varrho B\big(2\varrho\mathfrak{B_c}
+2e^{Ar^2}\big(-\varrho\mathfrak{B_c}+3\mathfrak{B_c}Br^2\big(r^2-2\varrho\big)-2\big)\\\nonumber
&+3B^2r^4(8\varrho\mathfrak{B_c}-1)+12\varrho\mathfrak{B_c}Br^2+4\big)\big\}\bigg]^{-1}\bigg[2\big(e^{Ar^2}\big(r^2-\varrho\big)+\big(B^2r^4\\\nonumber
&+3Br^2-A\big(Br^2+2\big)r^2+1\big)\varrho\big)\big(4e^{Ar^2}\big(r^2-\varrho\big)+\big(-A^2r^4+3B^2r^4\\\nonumber
&+17Br^2-A\big(10Br^2+21\big)r^2+4\big)\varrho\big)\big\{r^2\varrho\big(3B(22\varrho\mathfrak{B_c}+1)r^2+88\varrho\mathfrak{B_c}\\\nonumber
&-4e^{Ar^2}\big(3r^2+\varrho\big)\mathfrak{B_c}+14\big)A^3+\big(\varrho\big(3B^2(10\varrho\mathfrak{B_c}+9)r^4+2Br^2(31\\\nonumber
&-36\varrho\mathfrak{B_c})-8\varrho\mathfrak{B_c}+11\big)-4e^{Ar^2}\big(Br^4+(-B\varrho+5\mathfrak{B_c}\varrho+1)r^2+\varrho\\\nonumber
&\times(2\varrho\mathfrak{B_c}-1)\big)\big)A^2+\big(4e^{Ar^2}\big(B^2\big(r^2-\varrho\big)(\varrho\mathfrak{B_c}+1)r^2
-B\varrho+\varrho\mathfrak{B_c}-1\big)\\\nonumber
&-B\varrho\big(39B^2(2\varrho\mathfrak{B_c}+1)r^4+2B(64\varrho\mathfrak{B_c}+55)r^2+8\varrho\mathfrak{B_c}+69\big)\big)A+B\\\nonumber
&\big(e^{Ar^2}\big(4B\big(2r^2-\varrho\big)(\varrho\mathfrak{B_c}+1)+8\big)+B\varrho\big(9B^2(6\varrho\mathfrak{B_c}+1)r^4+2B(28\varrho\mathfrak{B_c}\\\nonumber
&+33)r^2+6(\varrho\mathfrak{B_c}+7)\big)\big)\big\}r^2-2\big(e^{Ar^2}\big(A\big(r^2-\varrho\big)+1\big)+\big(B\big(2Br^2+3\big)\\\nonumber
&-2A\big(Br^2+1\big)\big)\varrho\big)\big(4e^{Ar^2}\big(r^2-\varrho\big)+\big(-A^2r^4+3B^2r^4+17Br^2-A\\\nonumber
&\times\big(10Br^2+21\big)r^2+4\big)\varrho\big)\big\{A^3\varrho\big(B(22\varrho\mathfrak{B_c}+1)r^2+44\varrho\mathfrak{B_c}+7\big)r^4\\\nonumber
&+A^2\varrho\big(B^2(10\varrho\mathfrak{B_c}+9)r^4+B(31-36\varrho\mathfrak{B_c})r^2-8\varrho\mathfrak{B_c}-4\big(3r^2+\varrho\big)\\\nonumber
&\times\mathfrak{B_c}e^{Ar^2}+11\big)r^2+B\big(4e^{Ar^2}\big(r^2-\varrho\big)\big(B(\varrho\mathfrak{B_c}+1)r^2+2\big)+\varrho\big(3B^3r^6\\\nonumber
&\times(6\varrho\mathfrak{B_c}+1)+B^2(28\varrho\mathfrak{B_c}+33)r^4+6B(\varrho\mathfrak{B_c}+7)r^2+8\big)\big)-A\big(4e^{Ar^2}\\\nonumber
&\times\big(r^2-\varrho\big)\big(Br^2-\varrho\mathfrak{B_c}+1\big)+\varrho\big(13B^3(2\varrho\mathfrak{B_c}+1)r^6+B^2(64\varrho\mathfrak{B_c}+55)r^4\\\nonumber
&+B(8\varrho\mathfrak{B_c}+69)r^2-4\varrho\mathfrak{B_c}+4\big)\big)\big\}r^2-2\big(e^{Ar^2}\big(r^2-\varrho\big)+\big(B^2r^4\\\nonumber
&+3Br^2-A\big(Br^2+2\big)r^2+1\big)\varrho\big)\big(4e^{Ar^2}\big(A\big(r^2-\varrho\big)+1\big)+\big(-2A^2r^2\\\nonumber
&+B\big(6Br^2+17\big)-A\big(20Br^2+21\big)\big)\varrho\big)\big\{A^3\varrho\big(B(22\varrho\mathfrak{B_c}+1)r^2+7\\\nonumber
&+44\varrho\mathfrak{B_c}\big)r^4+A^2\varrho\big(B^2(10\varrho\mathfrak{B_c}+9)r^4+B(31-36\varrho\mathfrak{B_c})r^2-8\varrho\mathfrak{B_c}\\\nonumber
&-4e^{Ar^2}\big(3r^2+\varrho\big)\mathfrak{B_c}+11\big)r^2+B\big(4e^{Ar^2}\big(r^2-\varrho\big)\big(B(\varrho\mathfrak{B_c}+1)r^2+2\big)\\\nonumber
&+\varrho\big(3B^3(6\varrho\mathfrak{B_c}+1)r^6+B^2(28\varrho\mathfrak{B_c}+33)r^4+6B(\varrho\mathfrak{B_c}+7)r^2+8\big)\big)\\\nonumber
&-A\big(4e^{Ar^2}\big(r^2-\varrho\big)\big(Br^2-\varrho\mathfrak{B_c}+1\big)+\varrho\big(13B^3(2\varrho\mathfrak{B_c}+1)r^6+B^2\\\nonumber
&\times(64\varrho\mathfrak{B_c}+55)r^4+B(8\varrho\mathfrak{B_c}+69)r^2-4\varrho\mathfrak{B_c}+4\big)\big)\big\}r^2+2\big(e^{Ar^2}\big(r^2\\\nonumber
&-\varrho\big)+\big(B^2r^4+3Br^2-A\big(Br^2+2\big)r^2+1\big)\varrho\big)\big(4e^{Ar^2}\big(r^2-\varrho\big)+\big(4\\\nonumber
&-A^2r^4+3B^2r^4+17Br^2-A\big(10Br^2+21\big)r^2\big)\varrho\big)\big\{A^3\varrho\big(B(22\varrho\mathfrak{B_c}+1)\\\nonumber
&\times
r^2+44\varrho\mathfrak{B_c}+7\big)r^4+A^2\varrho\big(B^2(10\varrho\mathfrak{B_c}+9)r^4+B(31-36\varrho\mathfrak{B_c})r^2\\\nonumber
&-8\varrho\mathfrak{B_c}-4e^{Ar^2}\big(3r^2+\varrho\big)\mathfrak{B_c}+11\big)r^2+B\big(4e^{Ar^2}\big(r^2-\varrho\big)\big(B(\varrho\mathfrak{B_c}+1)\\\nonumber
&\times
r^2+2\big)+\varrho\big(3B^3(6\varrho\mathfrak{B_c}+1)r^6+B^2(28\varrho\mathfrak{B_c}+33)r^4+6B(\varrho\mathfrak{B_c}+7)\\\nonumber
&\times
r^2+8\big)\big)-A\big(4e^{Ar^2}\big(r^2-\varrho\big)\big(Br^2-\varrho\mathfrak{B_c}+1\big)+\varrho\big(13B^3(2\varrho\mathfrak{B_c}+1)\\\nonumber
&\times
r^6+B^2(64\varrho\mathfrak{B_c}+55)r^4+B(8\varrho\mathfrak{B_c}+69)r^2-4\varrho\mathfrak{B_c}+4\big)\big)\big\}-4\bigg]\bigg|.
\end{align}

\end{document}